\documentclass[a4paper]{jpconf}
\usepackage{amssymb}
\usepackage{amsmath}
\usepackage{graphicx}
\begin{document}
\title{Evaluating Reactor Antineutrino Signals for WATCHMAN}

\author{S.T. Dye}

\address{University of Hawaii at Manoa, 2505 Correa Road, Honolulu Hawaii 96822 U.S.A.}

\ead{sdye@phys.hawaii.edu}

\begin{abstract}
Increasing the distance from which an antineutrino detector is capable of monitoring the operation of a registered reactor, or discovering a clandestine reactor, strengthens the Non-Proliferation of Nuclear Weapons Treaty. This paper presents calculations of reactor antineutrino interactions from quasi-elastic neutrino-proton scattering and elastic neutrino-electron scattering in a water-based detector operated $\gtrsim10$ km from a commercial power reactor. It separately calculates signal from the proximal reactor and background from all other registered reactors. The main results are differential and integral interaction rates from the quasi-elastic and elastic processes. There are two underground facilities capable of hosting a detector ($\sim1$ kT H$_2$O) project nearby ($L\sim20$ km) an operating  commercial reactor ($P_{th}\sim3$ GW). These reactor-site combinations, which are under consideration for project WATCHMAN, are Perry-Morton on the southern shore of Lake Erie in the United States and Hartlepool-Boulby on the western shore of the North Sea in England. The signal rate from the proximal reactor is about five times greater at the Morton site than at the Boulby site due to shorter reactor-site separation distance, larger reactor thermal power, and greater neutrino oscillation survival probability. Although the background rate from all other reactors is larger at Morton than at Boulby, it is a smaller fraction of the signal rate from the proximal reactor at Morton than at Boulby. Moreover, the Hartlepool power plant has two cores whereas the Perry plant has a single core. The Boulby site, therefore, offers an opportunity for remotely monitoring the on/off cycle of a reactor core under more stringent conditions than does the Morton site.
\end{abstract}

\section{Introduction}
Monitoring the operation of a reactor from a remote location through the detection of antineutrinos is a nuclear nonproliferation goal \cite{adamb10} and the objective of project WATCHMAN \cite{watchman}. Nuclear monitoring activities and studies primarily utilize quasi-elastic antineutrino-proton scattering \cite{bowden09,snif10,nudar13}, commonly called the inverse beta decay (IBD) reaction. Coincidence counting of both products of the reaction $\overline{\nu}_e p \rightarrow e^+ n$, a positron and a neutron, is a traditional antineutrino detection technique. The positron signal estimates the antineutrino energy and the neutron tag reduces background. Neutrino oscillations distort the energy spectrum of detected antineutrinos, providing important information on the distance to the source \cite{dye09,nudar13}. While the angular distribution of the positrons is nearly isotropic in the energy range of reactor antineutrinos, the neutrons scatter in the forward direction \cite{vogel99}. Resolving the direction to the source of antineutrinos using inverse beta decay relies on measuring the direction of the outgoing neutron. Although the uncharged neutron does not directly produce a trail of ionization, the location of its capture is generally farther from the antineutrino source than the positron track. Asymmetry in the ensemble of positions of neutron capture relative to positron production is apparent in the data of several reactor antineutrino detection experiments, using scintillating liquid \cite{chooz00,paloverde00}. An additional reaction to employ for the remote monitoring of nuclear reactors is elastic neutrino-electron scattering (ES). This reaction $\overline{\nu}_l e^- \rightarrow \overline{\nu}_l e^-$ with $(l=e,\mu,\tau)$, whether induced by a neutrino or an antineutrino, always knocks the electron into the hemisphere directed away from the source. The scattered electron ionizes the detection medium and, if sufficiently energetic, produces Cherenkov radiation in a cone around the direction of the electron track. Reconstructing the ionization trail or the Cherenkov ring estimates track direction, which rejects background. Monitoring nuclear fusion in the Sun using directional Cherenkov radiation from elastic neutrino-electron scattering in water is well established. Adding information on reactor antineutrino direction to flux and spectral information enhances nuclear monitoring capabilities \cite{nudar13}. Water doped with gadolinium, which facilitates the detection of neutron captures, enhances the opportunity to combine flux and spectral information from quasi-elastic neutrino-proton scattering with directional information from elastic antineutrino-electron scattering.

\section{Calculations}
This paper presents calculations of reactor antineutrino interactions in the water target of a distant detector. These calculations are the foundation for assessing the capability of remotely monitoring the operation of a nuclear reactor. Scaling the results to detectors of different sizes, target media, and standoff distances is straightforward.

\subsection{Reactor-Site Combinations}
There are two unique combinations of a commercial reactor ($P_{th}\sim3$ GW) operating nearby ($L\sim20$ km) an underground facility capable of hosting a detector ($\sim1$ kT H$_2$O) project. These reactor-site combinations are Perry-Morton ($L=13$ km, $P_{th}=3758$ MW from 1 core) on the southern shore of Lake Erie in the United States and Hartlepool-Boulby ($L=25$ km, $P_{th}=3000$ MW from 2 cores) on the western shore of the North Sea in England. Both the Perry-Morton and the Hartlepool-Boulby combinations are under consideration for demonstrating remote monitoring of a nuclear reactor through the detection of antineutrinos \cite{watchman}. The overburden at Boulby ($D\sim2800$ m.w.e.) is greater than at Morton ($D\sim1500$ m.w.e.). Overburden is important for estimating cosmogenic background \cite{hellfeld}. Background sources are other reactors, solar neutrinos, geo-neutrinos, cosmogenic nuclides, neutrons, and detector noise, including radon contamination. This paper considers background only from other reactors. 

\subsection{Reactor Antineutrino Signal Spectrum}
Nuclear reactors generate heat by fissioning uranium and plutonium isotopes. The four main isotopes are $^{235}$U, $^{238}$U, $^{239}$Pu, and $^{241}$Pu. Antineutrinos emerge from the beta decay of the many fission fragments. An estimate of the energy spectrum of antineutrinos sums the exponential of a degree two polynomial of antineutrino energy ($E_\nu$) for each of the main isotopes \cite{vogel_engel}, $\lambda(E_\nu)=\mathrm{exp}(a_0+a_1E_\nu+a_2E_\nu^2)$. Each isotope ($i=1, 2,3,4$) contributes a fraction of the reactor power ($p_i$), releasing an average energy per fission ($Q_i$). The estimated energy spectrum of the reactor antineutrino emission rate is \cite{baldoncini}
\begin{equation}
dR/dE_\nu=P_{th}\sum_i \frac{p_i}{Q_i} \lambda_i(E_\nu).
\label{reacspec}
\end{equation}
Power fraction \cite{bellini10} and energy per fission \cite{ma13} values for each isotope are found in the literature. This paper assumes that boiling water reactors (Perry) and gas cooled reactors (Hartlepool) produce the same energy spectrum of antineutrinos.

\subsection{Neutrino Oscillations}
Neutrino flavors ($e$, $\mu$, $\tau$) are quantum mechanical mixtures of three neutrino mass states ($m_1$, $m_2$, $m_3$). Mixture varies with distance and energy, according to the well established phenomenon of neutrino oscillations. The probability that an electron antineutrino of energy $E_\nu$ loses its flavor after traveling a distance $L$ is $P_{e\rightarrow\mu,\tau}(L,E_\nu)$.
The complementary probability, $P_{e\rightarrow e}(L,E_\nu) = 1 - P_{e\rightarrow\mu,\tau}(L,E_\nu)$, gauges survival of electron flavor. This paper estimates the spectra of detectable reactor antineutrino interactions using oscillation parameter values \cite{capozzi16} from a recent global analysis and assuming normal mass ordering ($m_3>m_2>m_1$).

Reactor antineutrinos which oscillate to $\overline\nu_{\mu}$ and $\overline\nu_{\tau}$ do not have enough energy to initiate inverse beta decay. They do, however, interact by elastic scattering although with smaller cross section than $\overline\nu_e$. Neutrino oscillations reduce the interaction rate and distort the energy spectrum of the detected charged lepton more for IBD than for ES. The spectral distortion of IBD interactions provides important information on the distance to the source of antineutrinos \cite{dye09,nudar13}.

\subsection{Antineutrino Scattering}
Reactor antineutrinos scatter in ordinary matter primarily by two processes. The dominant reaction is IBD with cross section that follows from the V-A theory of weak interactions \cite{vogel99}. Calculations of the IBD cross section, which herein neglect energy-dependent recoil, weak magnetism, and radiative corrections, use the coefficient $\sigma^{IBD}_0 = 9.62 \times10^{-44}$ cm$^2$/MeV$^2$. IBD, which increases quadratically with neutrino energy, provides no sensitivity below $1.8$ MeV. Standard electroweak theory gives the cross section for the sub-dominant reaction ES \cite{fukuyana} with coefficient $\sigma^{ES}_0 = 1.436\times10^{-45} \mathrm{cm}^2/\mathrm{MeV}$. ES, which increases linearly with neutrino energy, has no threshold. The detectable particle in ES is an atomic electron, which always scatters forward relative to the direction of the neutrino. Reactor antineutrinos that oscillate to $\overline{\nu}_{\mu,\tau}$ still induce ES although with smaller interaction cross section than $\overline{\nu}_e$, thereby recovering some of the signal. The same $\overline{\nu}_{\mu,\tau}$ lack the energy to create a $\mu^+$ or $\tau^+$ and therefore do not induce quasi-elastic scattering. The resulting distortion of the energy spectrum detected by IBD reveals the full effect of neutrino oscillations.

\subsection{Interaction Rates}
The number $N(L,E_\nu)$ of reactor antineutrino interactions as a function of standoff distance $L$ and antineutrino energy $E_\nu$ at a given site follows from
\begin{equation}
N(L,E_\nu)=\frac{n\tau}{4\pi L^2}\int\sigma(E_\nu)\frac{dR}{dE_\nu}P(L,E_\nu)dE_\nu,
\end{equation}
where $n$ is the number of targets (free protons or hydrogen nuclei for IBD and atomic electrons for ES), $\tau$ is the exposure time, $\sigma(E_\nu)$ is the energy-dependent cross section, $dR/dE_\nu$ is as given in \eqref{reacspec}, and $P(L,E_\nu)$ is the oscillation survival probability. For the present study a convenient unit of exposure corresponds to a detector with a target mass of $1$ kT of water, where $n_p=6.69\times10^{31}$ for IBD and $n_e=3.35\times10^{32}$ for ES, that is operated for $1$ month, where $\tau=2.63\times10^6$ s.

The calculations here estimate 513 (111) IBD (ES) interactions per kT-month for Perry-Morton and 87 (20) IBD (ES) interactions per kT-month for Hartlepool-Boulby. These estimates apply to the proximal reactor cores operating continuously at full power and perfect detection efficiency. The Perry-Morton signal rates are more than five times greater than the Hartlepool-Boulby signal rates due to shorter reactor-site separation distance, larger reactor thermal power, and higher neutrino oscillation survival probability. Estimated interaction rates from other reactors are 25 (7) IBD (ES) per kT-month at Morton and 14 (4) IBD (ES) interactions per kT-month at Boulby.

\begin{figure}[h]
\begin{minipage}{17.5pc}
\includegraphics[trim = 5mm 40mm 15mm 40mm, clip, scale=0.45, width=17.5pc]{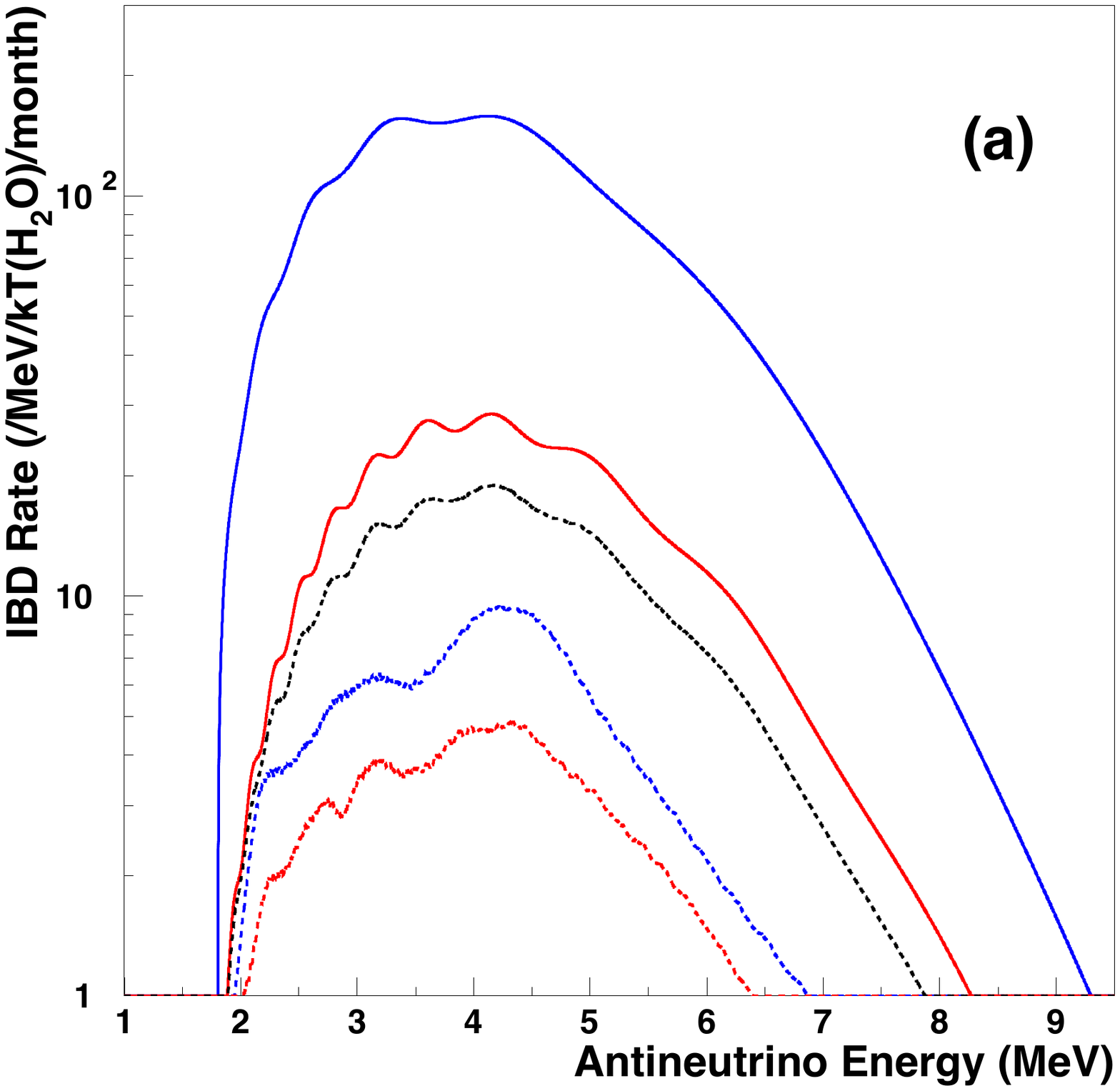}
\end{minipage}\hspace{2pc}
\begin{minipage}{17.5pc}
\includegraphics[trim = 5mm 40mm 15mm 40mm, clip, scale=0.45, width=17.5pc]{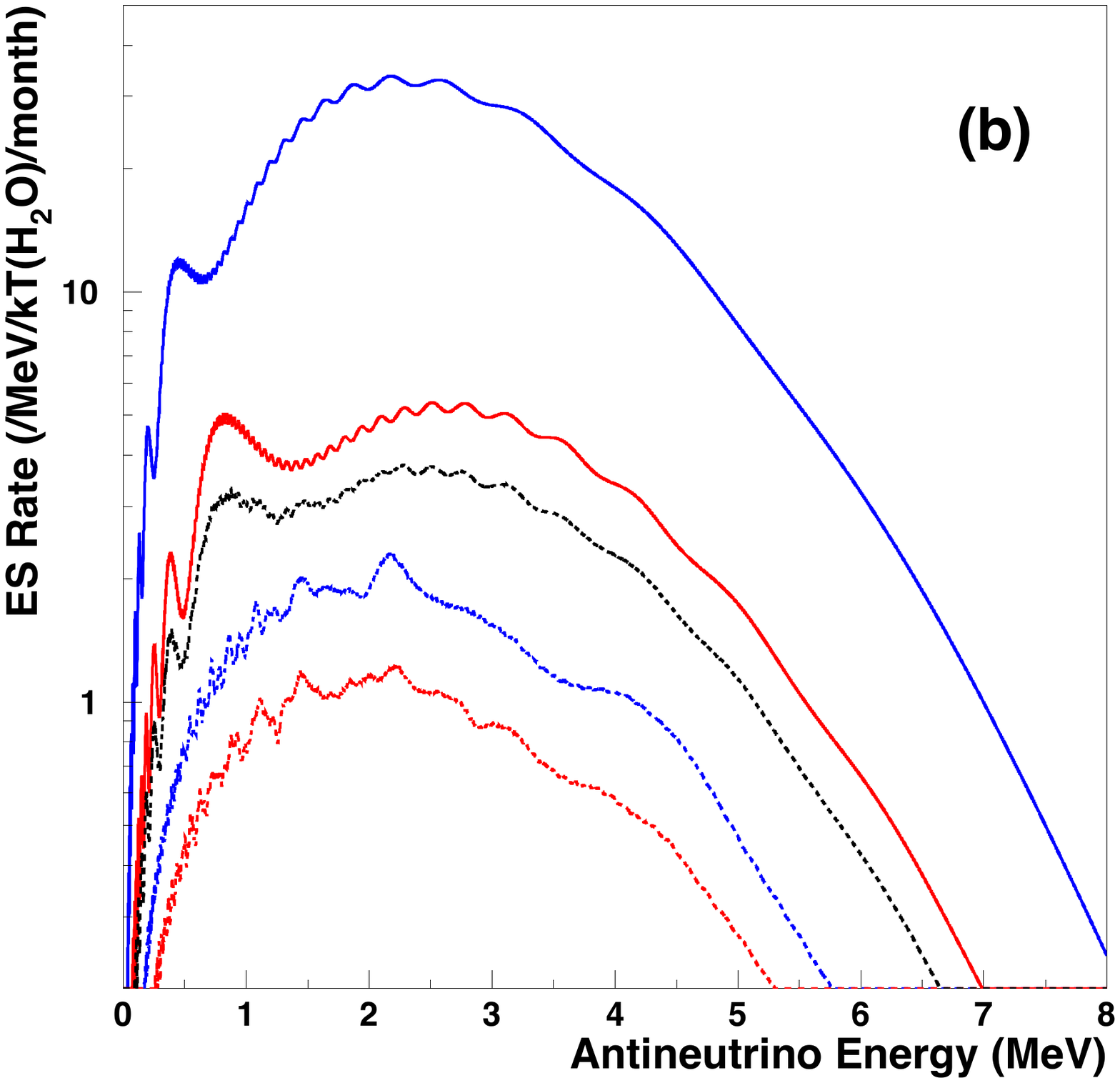}
\end{minipage} 
\end{figure}

\begin{figure}[h]
\begin{minipage}{17.5pc}
\includegraphics[trim = 5mm 40mm 15mm 40mm, clip, scale=0.45, width=17.5pc]{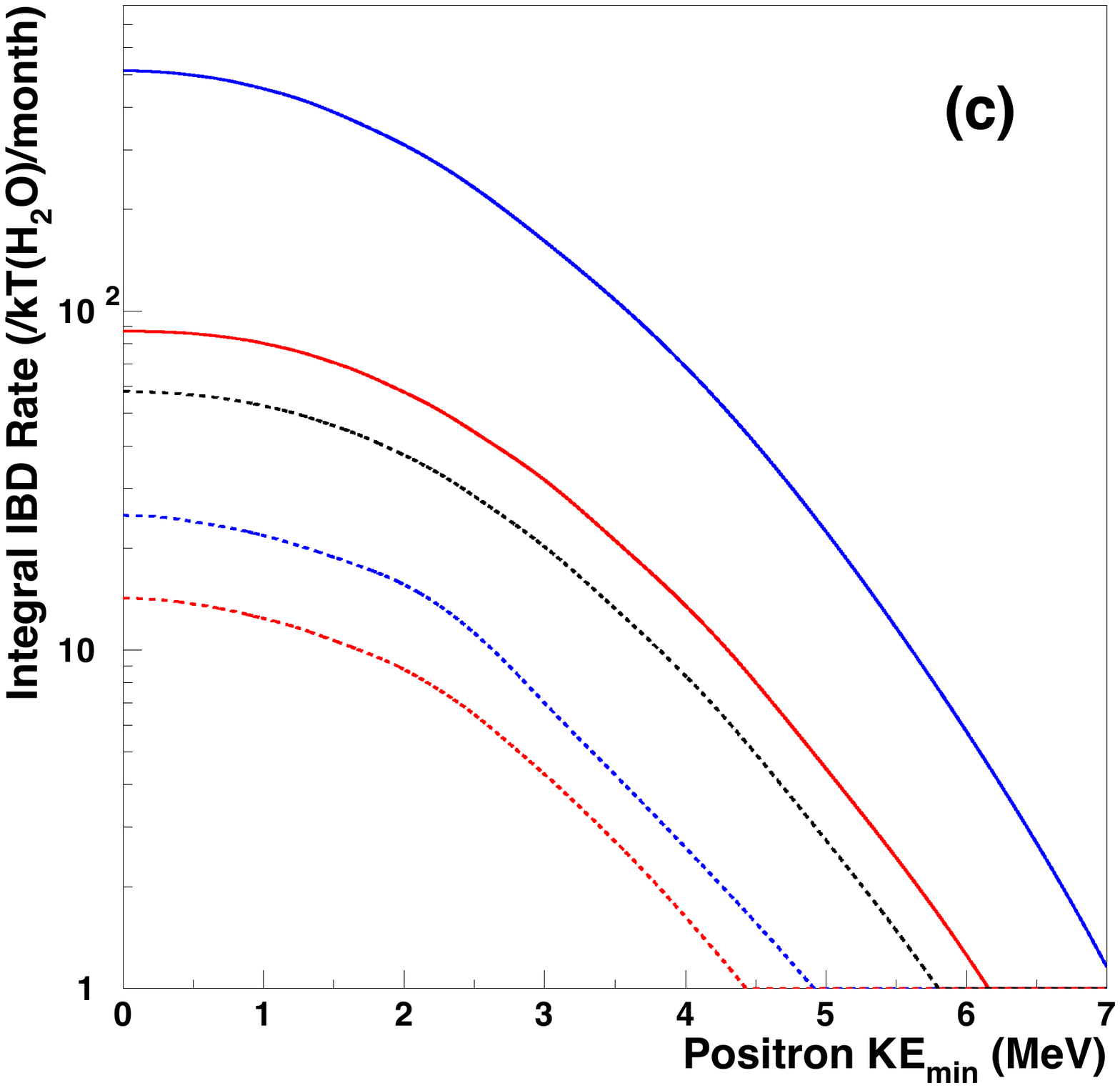}
\end{minipage}\hspace{2pc}
\begin{minipage}{17.5pc}
\includegraphics[trim = 5mm 40mm 15mm 40mm, clip, scale=0.45, width=17.5pc]{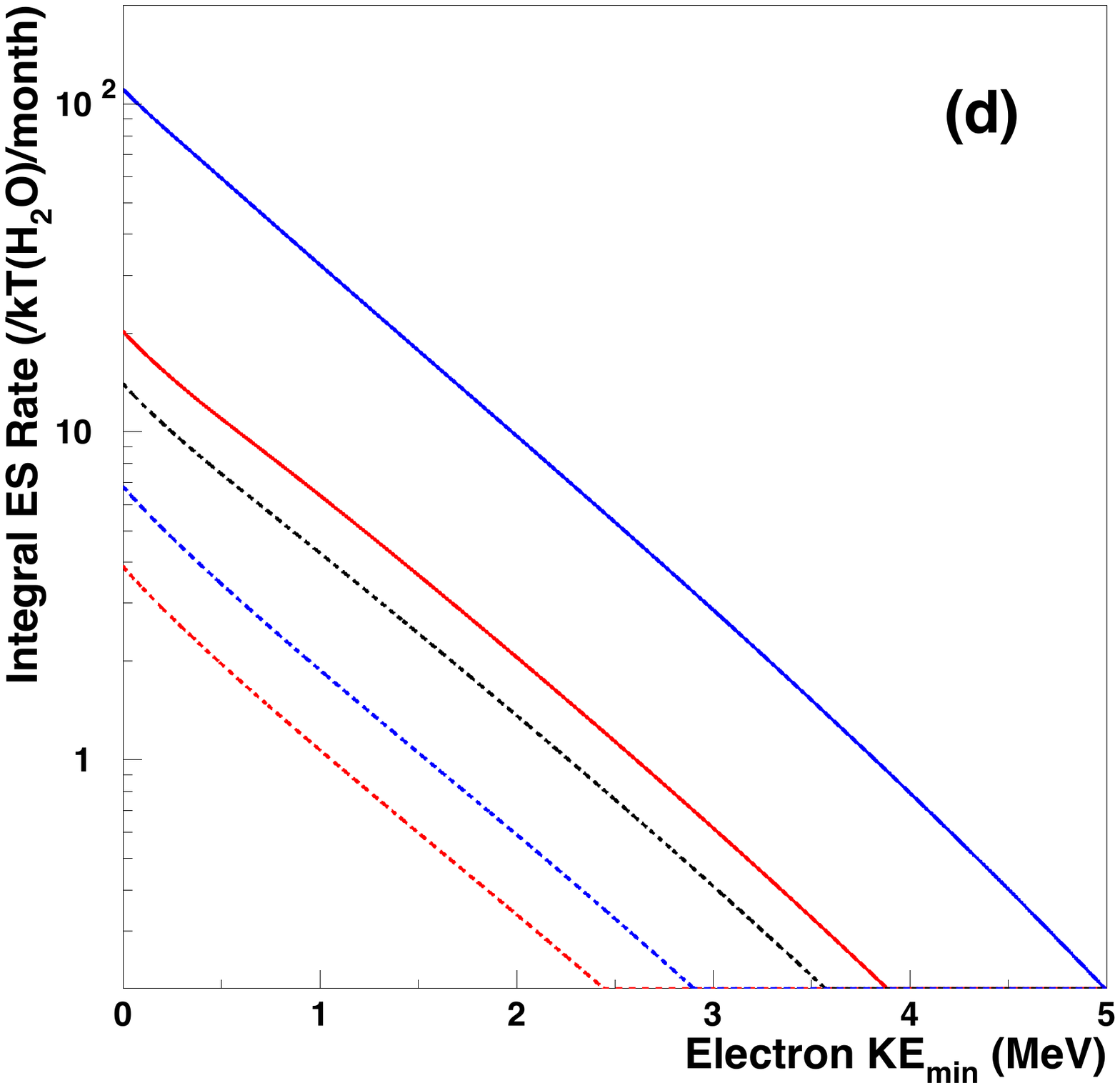}
\end{minipage} 
\caption{\label{fig:scat_rate} Rate of interactions as a function of reactor antineutrino energy from IBD {\bf (a)} and ES {\bf (b)}. Integral rate of interactions as a function of scattered lepton energy from IBD {\bf (c)} and ES {\bf (d)}. Solid blue (red) curves show the signal at the detection site from the proximal reactor operating at maximum power: Perry-Morton (Hartlepool-Boulby). Dashed blue (red) curves show the signal at the detection site from all other reactors operating at annual average power: Other-Morton (Other-Boulby). Dashed black curves show the signal at Boulby with one of the Hartlepool cores shut down.}
\end{figure}

The differential interaction rates as a function of antineutrino energy are shown at the two detector locations for IBD in figure~\ref{fig:scat_rate}(a) and for ES in figure~\ref{fig:scat_rate}(b). The integral interaction rates above a given kinetic energy of the scattered lepton are shown at the two detector locations for IBD in figure~\ref{fig:scat_rate}(c) and for ES in figure~\ref{fig:scat_rate}(d). All panels of figure~\ref{fig:scat_rate} show separately the signal due to the proximal reactor (Perry or Hartlepool) operating at full power and the background from all other reactors operating at their annual average load factor during the year 2014 \cite{infnsite}.  Because the Hartlepool reactor has two cores, the figures display the spectrum representing one-half the signal from Hartlepool added to the spectrum from all other reactors. This represents the background at Boulby for monitoring the on/off cycle of one of the Hartlepool reactor cores.

\section{Conclusions}
This paper presents calculations of the reactor antineutrino interactions, from both quasi-elastic neutrino-proton scattering (IBD) and elastic neutrino-electron scattering (ES), in a water-based detector operated a remote distance from a commercial power reactor. It separately calculates signal from the proximal reactor and background from all other registered reactors. Calculations are for two specific examples of a commercial reactor ($P_{th}\sim3$ GW) operating nearby ($L\sim20$ km) an underground facility capable of hosting a detector ($\sim1$ kT H$_2$O) project. These reactor-site combinations are Perry-Morton on the southern shore of Lake Erie in the United States and Hartlepool-Boulby on the western shore of the North Sea in England. Scaling the results to detectors of different sizes and target media is straightforward. The signal rate from the proximal reactor is about five times greater at the Morton site than at the Boulby site due to shorter reactor-site separation distance, larger reactor thermal power, and greater neutrino oscillation survival probability. Although the background rate from all other reactors is larger at Morton than at Boulby, it is a smaller fraction of the signal rate from the proximal reactor at Morton than at Boulby. Moreover, the Hartlepool power plant has two cores whereas the Perry plant has a single core. These conditions make monitoring the operation cycle of a nuclear reactor more challenging at the Boulby site than at the Morton site. The Boulby site, therefore, offers an opportunity for demonstrating remote reactor monitoring under more stringent conditions than does the Morton site.

\section*{Acknowledgments}
This work was supported in part by Lawrence Livermore National Laboratory.

\end{document}